\documentclass[letterpaper, 10 pt, conference]{IEEEconf}  % Comment this line out
\IEEEoverridecommandlockouts                              % This command is only
\usepackage{graphics} % for pdf, bitmapped graphics files
\usepackage{epsfig} % for postscript graphics files
\usepackage{mathptmx} % assumes new font selection scheme installed
\usepackage{times} % assumes new font selection scheme installed
\usepackage{amsmath} % assumes amsmath package installed
\usepackage{amssymb}  % assumes amsmath package installed
\usepackage{mathtools}
\usepackage{booktabs}
\usepackage{multirow}
\usepackage{xcolor}
\usepackage{hyperref}
\usepackage{caption}

\newtheorem{theorem}{Theorem}[section]

\newtheorem{lemma}[theorem]{Lemma}

\newtheorem{definition}[theorem]{Definition}
\newtheorem{assumption}[theorem]{Assumption}

\title{\LARGE \bf
State Space Models as Foundation Models:\\A Control Theoretic Overview
}

\author{Carmen Amo Alonso$^\ast$, Jerome Sieber$^\ast$, and Melanie N. Zeilinger% <-this % stops a space
\thanks{*\ \textbf{These authors contributed equally; ordered alphabetically.}}%
\thanks{All authors are with the Institute for Dynamic Systems and Control, ETH Zurich, 8092 Zurich, Switzerland
        {\tt\small \{camoalonso,jsieber,mzeilinger\}@ethz.ch}. This work was partially supported by the ETH AI Center.}%
}
\begin{document}

\maketitle
\thispagestyle{empty}
\pagestyle{empty}

%%%%%%%%%%%%%%%%%%%%%%%%%%%%%%%%%%%%%%%%%%%%%%%%%%%%%%%%%%%%%%%%%%%%%%%%%%%%%%%%
\begin{abstract}
In recent years, there has been a growing interest in integrating linear state-space models (SSM) in deep neural network architectures of foundation models. This is exemplified by the recent success of Mamba, showing better performance than the state-of-the-art Transformer architectures in language tasks. Foundation models, like e.g. GPT-4, aim to encode sequential data into a latent space in order to learn a compressed representation of the data. The same goal has been pursued by control theorists using SSMs to efficiently model dynamical systems. Therefore, SSMs can be naturally connected to deep sequence modeling, offering the opportunity to create synergies between the corresponding research areas. This paper is intended as a gentle introduction to SSM-based architectures for control theorists and summarizes the latest research developments. It provides a systematic review of the most successful SSM proposals and highlights their main features from a control theoretic perspective. Additionally, we present a comparative analysis of these models, evaluating their performance on a standardized benchmark designed for assessing a model's efficiency at learning long sequences.
\end{abstract}


%%%%%%%%%%%%%%%%%%%%%%%%%%%%%%%%%%%%%%%%%%%%%%%%%%%%%%%%%%%%%%%%%%%%%%%%%%%%%%%%
\section{Introduction}
Recently, foundation models have become central to the field of artificial intelligence. These models are large-scale learning models that are initially pretrained on extensive datasets, and subsequently fine-tuned for specific tasks. The term foundation models highlights these models' capability to learn and effectively generalize across a wide array of modalities, encompassing language, audio, images, video, genomics, and more. At their core, the predominant architecture for foundation models is the Transformer~\cite{Transformer}. This architecture, based on the attention mechanism, allows to efficiently process information and model global dependencies in complex data; but it suffers from two main limitations. One is computational complexity: it requires the complete sequence to be fed into the model every time an output is generated, which results in poor scalability with the time horizon window,\footnote{Referred to as \emph{input length} in Transformer's literature.} and therefore poor performance in long context tasks \cite{Tay2021}. The other limitation is explainability: despite its simple mathematical representation, it is currently not possible to interpret or understand the choice of outputs made by the Transformer \cite{bender2021dangers}. Efforts to address the scalability challenges of Transformers have led to various architectural variants that still leverage the merits of the attention mechanism. Examples of such variants are the Longformer~\cite{Longformer}, BigBird~\cite{Bigbird}, the Reformer~\cite{Reformer}, the Performer~\cite{Performer}, and approaches leveraging Axial Attention~\cite{Axialattention}. However, despite extensive research on these fronts, the proposed solutions often degrade the inherent merits of the architecture or fail to perform well in practice \cite{Tay2021}. 

A recent and promising research avenue proposes to fully replace the attention mechanism with a different representation based on State Space Models (SSM). The advantage of the SSM representation lies in its recurrent nature, where only the latest input has to be passed to the model since the state is able to capture information about past inputs. Moreover, due to their mathematical structure, they are amenable to computationally efficient training and inference--in contrast to their predecessors, recurrent neural networks (RNNs)~\cite{Goodfellow2016}. This new family of SSM-based architectures has been shown to beat Transformers in long-context tasks such as the Long Range Arena (LRA) benchmark~\cite{Tay2021}, and recent proposals such as Mamba~\cite{mamba} exhibit performance and computational efficiency superior to state-of-the-art Transformers on long-context tasks. These results highlight the potential of SSMs to overcome many of the current limitations of Transformers. Although SSMs show great promise to serve as foundation models, most of the existing literature on SSMs focuses on providing performant architectures and efficient implementations. Despite the clear connection with control theory, in particular linear systems theory, to date a principled understanding of these models is lacking, and most design choices are motivated from an empirical performance rather than a systematic system theoretical viewpoint. There is large potential in leveraging existing system theoretic results and analysis to complement current implementations and enhance explainability, design and performance.

Towards this goal, the aim of this paper is to provide an overview of state-of-the-art SSMs from a control theoretical perspective. In Section~\ref{sec:ssms}, we provide an overview of the essential components and considerations in SSMs. In Section~\ref{sec:models}, we review the most relevant SSM proposals to date. Since these models were primarily motivated by their ability to handle long contexts, we present the first performance comparison to date on the LRA benchmark in Section~\ref{sec:simulations}. Lastly, we end in Section~\ref{sec:conclusion} with concluding remarks and open research questions that could help advance SSMs and cross-pollinate the fields of foundation models and systems and control theory. 

\section{State Space Models} \label{sec:ssms}
We first present a generic language modelling task to define the learning goal of a foundation model. Then, we give an overview of the state space model architecture, mathematical structure, and computational considerations that guide the SSMs introduced in the literature.

\subsection{Learning setup}
A foundation model, such as those used in language modeling, can be seen as a map between input and output signals, i.e.,
\begin{equation}\label{eqn:general-model}
y(k) = f(u(k),\dots,u(k-T);\theta),
\end{equation}
where at each time $k$, the output $y(k)$ is produced after evaluating an input signal of length $k-T$, i.e., $u(k),\dots,u(k-T)$, and a set of parameters $\theta$. The parameters $\theta$ are task dependent, and can be fine-tuned accordingly. Since the search space of general $f(\cdot;\theta)$ is too broad, different parameterizations of $f(\cdot; \theta)$ can be used to render the problem tractable. For instance, the model $f(\cdot; \theta)$ can consist of multiple stacked models like e.g. the Transformer or more recently SSMs. The architectural choice of $f(\cdot; \theta)$ is a fundamental factor in determining the success of the model at effectively learning structure from data.

The goal of a foundation model used as large language model is to learn a compressed representation of structure present in language in order to perform tasks like machine translation or human-level conversations (e.g. ChatGPT). To learn such a representation the parameterized model $f(\cdot; \theta)$ is presented with input-output pairs $(u(k),y(k))\ \forall k$, where $\theta$ represents the parameters. The parameters $\theta$ are then iteratively updated to minimize a loss function $\mathcal{L}(\cdot)$, i.e., iteratively solving the following optimization problem
\begin{equation}\label{eqn:model}
    \min_{\theta}\; \mathcal{L}(y - f(u; \theta)).
\end{equation}
For a language model the inputs $u$ are tokenized\footnote{An input token is the unit that represent the smallest meaningful components of the input data, whether it's text, images, or any other form of information that the model processes.} sentences and the outputs $y$ are a shifted version of the same inputs, i.e., an auto-regressive setup.

\subsection{Parametrization}
Let us consider the following \textit{continuous-time} linear system with dynamics
\begin{subequations} \label{eqn:dynamics_continuous}
\begin{align}
    \dot{x}(t) &= Ax(t) + Bu(t), \\
    y(t) &= Cx(t) + Du(t),
\end{align}
\end{subequations}
where $x\in\mathbb C^p$ represents the  complex-valued state, $u,\ y\in\mathbb R^q$ are the input and the output, respectively, and $t$ denotes the continuous-time index. We note that the input fed into the system denoted as $u$ is \emph{not} a \emph{control} input; it is seen as an exogenous input exciting the system \eqref{eqn:dynamics_continuous}. This choice of notation is made to maintain consistency with the corresponding literature. $A,\ B,\ C,\ D$ are complex-valued matrices of appropriate dimensions and in representation \eqref{eqn:dynamics_continuous}, these matrices are assumed to be time-invariant. When considering their time-varying version, a time sub-index would be appended, i.e., $A_t,\ B_t,\ C_t,\ D_t$. 

In the SSM literature, system \eqref{eqn:dynamics_continuous} is used as a black-box representation in a foundation model. Here, the exogenous input $u(t)$ represents a signal or input token fed into the model at a given time $t$. The state $x(t)$ represents the hidden state that stores the relevant information about the current and previous inputs up to time $t$, and $y(t)$ is the output of the model at time $t$. In a learning setup, the matrices $A,\ B,\ C,\ D$ are parameters, which are commonly learned via stochastic gradient descent. Since computational efficiency and initialization are essential aspects in this framework, the dynamic matrix $A$ is often assumed to have a particular structure. As such, SSMs are often referred to as \emph{Structured} SSMs.

\begin{assumption} \label{assump:diagonal}
    The dynamic matrix in dynamics \eqref{eqn:dynamics_continuous} has a diagonal structure, i.e., $A = diag(\lambda_1,\dots,\lambda_p)$ with $\lambda_i\in\mathbb C\ \forall i$. 
\end{assumption}

Although initial proposals \cite{Gu2021, Gu2022} deviate slightly from Assumption \ref{assump:diagonal}, most of the Structured SSMs literature assumes a diagonal $A$ matrix. Specific choices will be discussed in Section \ref{sec:models}.

\subsection{Discretization}
In order to implement a SSM, a discrete-time version of system \eqref{eqn:dynamics_continuous} is used. Hence, the implementation of system \eqref{eqn:dynamics_continuous} in discrete-time is
\begin{subequations} \label{eqn:dynamics_discrete}
\begin{align}
    x(k+1) &= \bar{A}x(k) + \bar{B}u(k), \\
    y(k) &= \bar{C}x(k) + \bar{D}u(k),
\end{align}
\end{subequations}
where $\bar A,\ \bar B,\ \bar C,\ \bar D$ are the discrete-time dynamic matrices discretized with time-step $\Delta\in\mathbb R$, possibly with complex-valued components, and $k$ denotes the discrete-time index. The choice of discretization scheme chosen varies widely among the proposed models in the SSM literature, and an overview is presented in Section~\ref{sec:models}.

We note that it is also possible to directly start from a discrete-time model as in equation \eqref{eqn:dynamics_discrete}, oblivious to its continuous-time representation \eqref{eqn:dynamics_continuous}. However, in most of the SSM literature, a continuous-time view of the dynamics is preferred in order to better motivate the choice of initialization for the dynamical matrices\cite{Gu2020}. 

\subsection{Structure and Initialization}
Since the dynamics \eqref{eqn:dynamics_continuous} are being learned via gradient descent, initialization of the parameters was found to be of crucial importance. In particular, the initial values of matrix $A$ have a significant impact on the performance after training: on a simple classification task, performance increases from 67\% when $A$ is randomly initialized, to 80\% when $A$ is initialized using a principled strategy \cite[Section~4.4]{Gu2022}. Different strategies and parametrizations have been proposed in order to achieve a successful initialization, i.e. an initialization that results in the state $x(k)$ being able to capture the recent history of the inputs $u(k),\dots,u(k-T)$ for some time horizon $T$. This property is referred to as \emph{memory} in the standard SSM literature. As is well-known in control theory, the memory of system \eqref{eqn:dynamics_discrete} is directly linked to the eigenvalues of matrix $A$. 

\begin{lemma} \label{lemm:eigenvalues}
    (Informal)
    A dynamical system with dynamics~\eqref{eqn:dynamics_discrete} has long-range memory, i.e., captures information from past inputs, if the eigenvalues of $A$ are inside the unit circle and very close to the unit circumference, i.e. $\vert eig(A) \vert \leq 1$ and $\vert eig(A) \vert \approx 1$ $\forall eig(A)$.
\end{lemma}

% \iffalse
% \begin{proof}
%     We start by writing the output at time $k$, $y(k)$ as a function of the inputs $u(k),\dots,u(0)$ as
%     \begin{align*}
%         y(k) &= \sum_{\tau=0}^k \bar C \bar A^{k-\tau} \bar B u(\tau)\\
%              &= \sum_{\tau=0}^k \bar C\ diag(\lambda_1,\dots,\lambda_p)^{k-\tau} \ \bar B u(\tau),
%     \end{align*}
%     where the second equality comes from Assumption \ref{assump:diagonal}. Hence, the ability of past inputs $u(k),\dots,u(0)$ to influence $y(k)$ is driven by the values of $\lambda_1,\dots,\lambda_p$. If $\lambda_i = 1$, then $\lambda_i^{n} = 1\  \forall n$. However, if $\lambda_i < 1$, then $ \lim_{n\rightarrow n} \lambda_i^{n} = 0$, where the rate with which $\lambda_i^{n}$ approaches $0$ with increased $n$ (forgetting past inputs) is driven by how close $\lambda_i$ is to $1$. By a similar argument, if $\lambda_i > 1$, then $ \lim_{n\rightarrow n} \lambda_i^{n} = \infty$, which would make it impossible to process current inputs due to instability. \qedsymbol
% \end{proof}
% \fi

Hence, the various initialization schemes presented in the SSM literature aim to ensure that the modulo of the eigenvalues of the learned $A$ matrix is approximately equal to (but not bigger than) $1$. For the initialization of the other matrices, i.e., $B, \, C$, and $D$, standard initialization methods are used, e.g., Glorot~\cite{Glorot2010} or LeCun~\cite{Klambauer2017}, which essentially draw the initial values from a transformed uniform or normal distribution. Therefore, we omit the initialization details of $B, \, C$, and $D$ in the following and refer the reader to the original papers \cite{Glorot2010,Klambauer2017}.

\subsection{Implementation}
One of the major challenges addressed in the SSM literature is how to efficiently learn (training time) and deploy (inference time) the recurrence \eqref{eqn:dynamics_discrete}. 

At inference time, a causal representation is needed since the model does not have access to excitation inputs beyond the current time step. For this reason, the recurrent representation \eqref{eqn:dynamics_discrete} is directly used starting with an initial excitation $u(1)$ and zero initial state $x(1)=0$. In order to speed up this process, \emph{parallel scans} algorithms~\cite{Blelloch1990} are used that efficiently compute the recurrence by computing each output component in parallel and caching intermediate results.

During training, it is possible (and desirable) to use a non-causal representation since input-output pairs $(u(k),y(k))$ are available for all $k$. Different techniques have been proposed in the literature. Some of the architectures can take advantage of parallel scan algorithms and use the recurrent representation from equation \eqref{eqn:dynamics_discrete}. Some other architectures rely on the convolutional representation of system \eqref{eqn:dynamics_discrete}, i.e.,
\begin{equation}\label{eqn:convolution}
    y(k) = \sum_{\tau=0}^k \bar C \bar A^{k-\tau} \bar B u(\tau).
\end{equation}
This convolutional representation allows for faster learning because the complete input sequence $u(k)\ \forall k$ can be passed through the model in one step.

In terms of learning algorithms, SSM models are commonly trained using a standard stochastic gradient descent variation, i.e. Adam~\cite{Kingma2014}, and  backpropagation~\cite{Goodfellow2016}. Additionally, they can utilize the same heuristic methods to improve training as other deep-learning models, e.g., dropout or normalization~\cite{Goodfellow2016}.

\begin{figure}[ht]
\centering
\includegraphics[width=\linewidth,trim={0 2mm 0 0},clip]{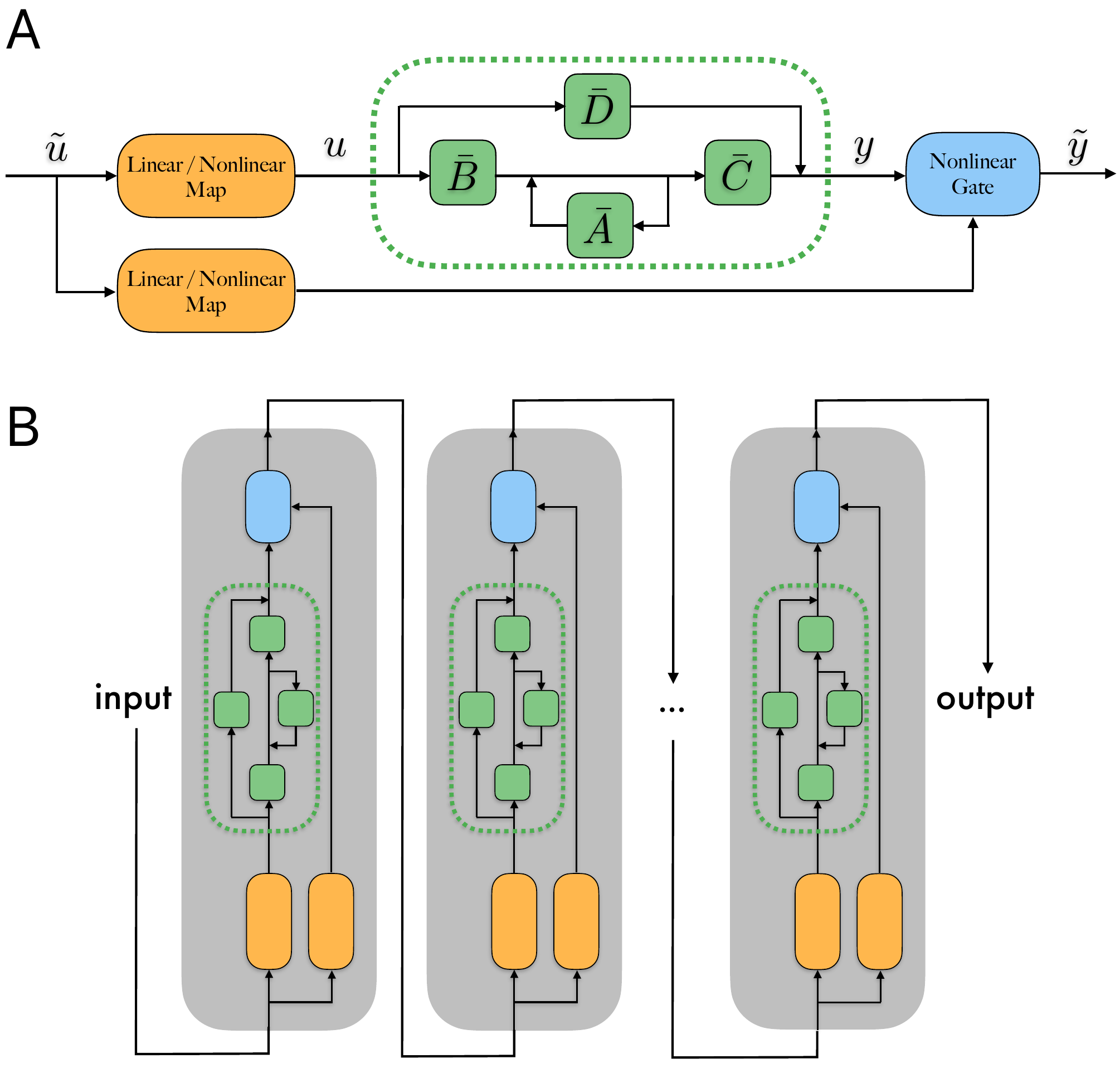}
    \caption{\textbf{A. General scaffolding of a SSM.}  The dynamical model \eqref{eqn:dynamics_discrete} is represented in green. The input to the SSM is pre-processed and forked off in a skip connection (lower signal). The nature of the pre-processing map (linear or nonlinear) depends on the specific scaffolding. The output of the recursion is then post-processed with a nonlinear gate.\\
    \textbf{B. Overall architecture of a SSM.} Each of the SSMs including its scaffolding (Fig. 1.A.) is structured in a layered fashion, where the output from one layer is the input to the next.}
    \label{fig:scaffolding} 
    \vspace{-12pt}
\end{figure}

\subsection{Scaffolding and Layers}
Although learning the dynamics in equation \eqref{eqn:dynamics_continuous} is a major focus of SSMs, these dynamics are not simply implemented in isolation. In fact, pre-processing of the input $u$ and post-processing of the output $y$ is necessary to ensure good performance. In this paper, we refer to the algebraic operations of pre- and post-processing as the \emph{scaffolding} surrounding the SSM computation in dynamics \eqref{eqn:dynamics_discrete}. A general overview of the architecture used in SSMs is provided in Figure \ref{fig:scaffolding}.

A collection of different scaffolding choices have been proposed in the literature, ranging from standard multilayer perceptron (MLP) choices to gating operations, as defined in Definition~\ref{def:gating}. In general, a linear or nonlinear map is performed on the input $\bar u$ before it is fed into system \eqref{eqn:dynamics_discrete}. Once the output $y$ has been computed, a gating operation is generally performed to control the flow of information from the input $\tilde u$ to the output $\tilde y$. Intuitively, the gate~$g(\tilde y, \tilde u)$ controls which outputs $\tilde y$ are set to zero based on the inputs $\tilde u$ via the softmax operation.

\begin{definition}\label{def:gating}
    Given two vectors $x_1,x_2\in\mathbb R^p$, a \emph{gating operation} is defined as $g(x_1,x_2):= x_1 \odot \sigma(Wx_2)$, where $W\in\mathbb R^{p\times p}$, $\odot$ is the element-wise multiplication, and $\sigma$ is the softmax operation.\footnote{In the SSM literature, sometimes other nonlinearities are used, such as ReLU, SiLU, etc.}
\end{definition}

As is common practice in deep learning, several layers of SSMs (dynamics \eqref{eqn:dynamics_continuous} and accompanying scaffolding) are stacked together, where each of them processes the output of the previous layer as its input, which is then fed into the next layer. This is possible since input $y$ and output $u$ are of the same dimension $\mathbb R^q$. For example on smaller tasks like e.g. the LRA benchmark~\cite{Tay2021}, a SSM is composed of 6 structurally-identical layers (with different dynamic matrices), and the size of the systems ranges in $p\in [64, 512],\ q\in[32, 1024]$. For language modelling the number of layers and system size can be significantly larger.

It is important to note that the choice and design of the scaffolding is not well-understood, and often the one that is most performant in practice is selected. 

\section{Review of existing methods} \label{sec:models}
In this section, we present an overview of the most prominent SSM proposals in the literature. Since existing SSMs build on each other, the order of presentation in this section is chronological. We provide details as to how each of the architectures tackles the considerations described in Section \ref{sec:ssms}. We also provide a summary of their main characteristics in Table \ref{table:ssms}.

\begin{figure*}
        \centering
        \begin{tabular}{@{}lcccccc@{}}
        \toprule
        \multirow{2}{*}{\textbf{Model}} & \multicolumn{5}{c}{\textbf{Features}} \\ \cmidrule(lr){2-6}
        & Parametrization & Discretization & Structure & Implementation & Scaffolding \\ \midrule
        S4~\cite{Gu2022} & LTI  & Bilinear & SISO & Convolution and Recurrence & MLP / H3 \\
        S4D~\cite{Gu2022s4d} & LTI & Exact & SISO & Convolution and Recurrence & MLP / H3 \\
        S5~\cite{Smith2023} & LTI  & Exact / Bilinear & MIMO & Parallel Scan & MLP / H3 \\
        LRU~\cite{Orvieto2023} & LTI  & None & MIMO & Parallel Scan & MLP / H3 \\ 
        S6~\cite{mamba} & LTV & Exact & MIMO & Custom Parallel Scan & Mamba \\
        RG-LRU~\cite{griffin} & LTV & None & MIMO & Custom Parallel Scan & Mamba / Hawk / Griffin \\
        \bottomrule 
        \end{tabular}
        \captionof{table}{Overview of the model features for the different SSM models considered. Accronyms used are as follows: Linear Time-Invariant (LTI), Linear Time-Varying (LTV), Single Input Single Output (SISO), Multiple Input Multiple Output (MIMO). Details on the scaffolding can be found in MLP \cite{MLP}, H3 \cite{Gu2020}, Mamba \cite{mamba}, Hawk and Griffin \cite{griffin}.}\label{table:ssms}
        \vspace{-12pt}
\end{figure*}

%\textbf{[Jerome: add in each of the architectures what they are good at (according to simulations). We need to mention, for instance, that S4 is better than Mamba in LRA, etc.]}

\subsection{Structured State Space Sequence Model (S4)}
The S4 model \cite{Gu2022} was the first proposed model based on a state space representation. 

\paragraph{Parametrization} The S4 model starts from a continuous time model~\eqref{eqn:dynamics_continuous}, where the structure imposed on matrix $A$ is 
\begin{equation}\label{s4:parametrization}
    A = diag(\lambda_1,\dots,\lambda_p) + rs^\star
\end{equation} 
with $\lambda_i\in\mathbb C\ \forall i$, and $r, \, s \in \mathbb{C}^p$. This is, a diagonal matrix plus a low-rank update. We note that this structure resembles a closed-loop dynamics matrix $A_{CL} = A + BK$.

\paragraph{Discretization} The discrete-time version \eqref{eqn:dynamics_discrete} is computed by applying the bilinear transform to dynamics \eqref{eqn:dynamics_continuous} with discretization step $\Delta\in\mathbb R$, i.e.,
\begin{equation}\label{eqn:s4_discretization}
    \bar{A} = (I - \frac{\Delta}{2}A)^{-1}(I + \frac{\Delta}{2}A), \qquad \bar{B} = (I - \frac{\Delta}{2}A)^{-1}\Delta B,
\end{equation}
$\bar{C} = C$ and $\bar{D} = D$. Note that this choice of discretization method couples the parameterizations of $\bar A$ and $\bar B$ via the discretization step $\Delta$, which is a common feature of most SSMs.

\iffalse
\begin{alignat*}{2}\label{s4:dynamics}
    \bar{A} &= (I - \frac{\Delta}{2}A)^{-1}(I + \frac{\Delta}{2}A), \qquad && \bar{C} = C \\
    \bar{B} &= (I - \frac{\Delta}{2}A)^{-1}\Delta B, \qquad && \bar{D} = D.
\end{alignat*}
 \fi 
 
\paragraph{Structure and Initialization}
The model is structured in a single input single output (SISO) manner, i.e., each component of the input (referred to as \emph{input channel}) $u_i$ for $i=1,\dots,q$ is fed into a separate system~\eqref{eqn:dynamics_discrete}, each producing a scalar output $y_j$ with $j=1,\dots,q$. Each dynamics matrix $A$ for each of the $q$ SISO subsystems is initialized using HiPPO theory \cite{Gu2020},  resulting in the eigenvalues shown in Figure~\ref{fig:init}. In essence, the HiPPO theory provides a mathematically grounded way to place the eigenvalues of a continuous-time dynamics matrix such that it can compress information over long input sequences into its state. Although the original S4 does not bias the initialization towards marginal stability to ensure long-range memory (as per Lemma \ref{lemm:eigenvalues}), the follow up work SaShiMi \cite{Goel2022} enforces $Re(\lambda_i)\in\mathbb R^{-}\ \forall i$ to ensure stability.

\paragraph{Implementation} At training time, a convolutional representation \eqref{eqn:convolution} is used. For efficient computation, the structure of $\bar A$~\eqref{s4:parametrization} is exploited since the Sherman-Morrison formula \cite{Sherman1950} can be used to compute its inverse in~\eqref{eqn:s4_discretization}, resulting in only the inversion of scalars. At inference time, the recurrent representation of the model \eqref{eqn:dynamics_discrete} is directly used. 

\paragraph{Scaffolding} Initially, the scaffolding proposed for the pre- and post-processing of the S4 block was identical to the one used for gated MLPs. Later on, a more sophisticated scaffolding, \emph{H3}~\cite{Fu2023}, was introduced to mimic the operations of a Transformer. The H3 scaffolding uses the sum of the original signal with a time-shifted version of the input signal for the linear map of the upper signal and a standard linear map for the lower signal in Figure~\ref{fig:scaffolding}.A. The post-processing remains a gating function.

\subsection{Diagonal Structured State Space Sequence Model (S4D)}
The initially proposed Diagonal State Space (DSS) \cite{Gupta2022} model and its enhancement S4D \cite{Gu2022s4d} build upon the S4 model. They simplify the structure of the dynamics matrices by introducing for the first time Assumption \ref{assump:diagonal}, which results in computational improvements.

\paragraph{Parametrization} The main contribution of the S4D paper is the introduction of a new, more efficient, structure of matrix $A$ consistent with Assumption \ref{assump:diagonal}: 
\begin{equation}\label{eqn:s4d_parametrization}
    A = diag(\lambda_1,\dots,\lambda_p).
\end{equation} 

\paragraph{Discretization} The discrete-time version \eqref{eqn:dynamics_discrete} is computed by applying exact discretization to dynamics \eqref{eqn:dynamics_continuous} with discretization step $\Delta\in\mathbb R$, i.e.,
\begin{equation}\label{eqn:s4d_discretization}
    \bar{A} = e^{\Delta A}, \qquad \bar{B} = (\Delta A)^{-1} (\bar{A} - I) \Delta B,
\end{equation}
$\bar{C} = C$ and $\bar{D} = D$.

\paragraph{Structure and Initialization} The SISO structure used in S4D is the same one as in S4. Initalization of S4D is also done using HiPPO theory, with the added insight that the resulting matrix can be diagonalized for added computational efficiency. Similar to SaShiMi~\cite{Goel2022}, the eigenvalues of $A$ used for initialization are constrained to lie in the negative halfplane. This initialization results in the eigenvalues shown in Figure~\ref{fig:init}.

\paragraph{Implementation} Similar to S4, a convolutional representation \eqref{eqn:convolution} is used at training time, and a recurrent representation \eqref{eqn:dynamics_discrete} at inference time. Given the diagonal structure of matrix $\bar A$, discretization \eqref{eqn:s4d_discretization} can be computed efficiently.

\paragraph{Scaffolding} The scaffolding of S4D is identical to the one used in S4. 

\subsection{Simplified Structured State Space Sequence Model (S5)}
The S5 parametrization~\cite{Smith2023} presents a simplification of the previously proposed S4D and leverages the concept of multiple input multiple output (MIMO) systems (as opposed to SISO) to simplify the architectural components and enhance computation.

\paragraph{Parametrization} The parametrization used is identical to S4D.

\paragraph{Discretization} S5 is amenable to both discretizations proposed in S4 and S4D: bilinear \eqref{eqn:s4_discretization} and exact \eqref{eqn:s4d_discretization}.

\paragraph{Structure and Initialization} The main contribution of the S5 model is the introduction of a MIMO interpretation of the previously proposed models, which leads to significant computational enhancements. In particular, the full input vector~$u \in \mathbb{R}^q$ is fed into a single MIMO system \eqref{eqn:dynamics_discrete} (of bigger dimension) as opposed to $q$ SISO scalar subsystems (of smaller dimension). This is achieved by stacking the subsystem matrices $\bar{A}, \, \bar{B}, \, \bar{C}$ used in S4 and S4D. The matrix $A$ is again initialized using HiPPO theory and results in the same initial eigenvalues as S4D (Figure~\ref{fig:init}).

\paragraph{Implementation} The MIMO structure together with the diagonal parameterization of $\bar{A}$ allows for parallel computation of the individual output components from the input components via a parallel scan algorithm~\cite{Blelloch1990}. As a result, both the computation at training time and the computation at inference time can be computed efficiently in their recurrent representation \eqref{eqn:dynamics_discrete}.

%a prefix sum is a mathematical relation of the form
%\begin{equation*}
%    y_k = \sum_{j=0}^{k-1} \alpha_{j} u_{j},
%\end{equation*}
%whose parallel implementation -- given some structure on $\alpha_k$ -- has been well studied in literature~\cite{Blelloch1990}.

\paragraph{Scaffolding} The MIMO representation allows for a simplication of the scaffolding previously proposed for S4 and S4D. The reason for this is that, although the stacked dynamics matrix $\bar{A}$ is diagonal, stacked matrices $\bar{B}, \, \bar{C}$ are dense and therefore couple the input and output components. This allows to remove a mixing layer present in the post-processing of the S4 and S4D output. 

\subsection{Linear Recurrent Unit (LRU)}
The LRU model attempts to simplify previous SSM proposals by unveiling their essential components. One of the main contributions of LRU is to explicitly encode long-range memory through eigenvalues. This allows to move away from the HiPPO theory and directly use a discrete-time model together with concepts of marginal stability from control theory.

\paragraph{Parameterization} The LRU model directly parameterizes the discrete-time dynamics~\eqref{eqn:dynamics_discrete}, i.e., 
\begin{equation}\label{eqn: lru_parameterization}
    \bar{A} = e^{-e^{\, diag(\lambda_1,\dots,\lambda_p)} + i\, diag(\theta_1,\dots,\theta_p)}, \qquad \bar{B} = e^{\gamma} \Gamma
\end{equation}
with $i$ the complex unit, $\lambda_j,\theta_j\in\mathbb R\ \forall j=1,\dots,p$, $\Gamma \in \mathbb{C}^{p \times q}$ a dense complex-valued matrix, and $\gamma \in \mathbb{R}$. Notice that this parameterization directly represents the diagonal entries of $\bar{A}$, and therefore the eigenvalues in polar coordinates, i.e. $a_j =  r_j + i\ \theta_j$ where $r_j = e^{-e^{\lambda_j}}$, is constrained to the interval~$[0,1]$ by construction. This is also the first parameterization that does not have shared parameters between $\bar{A}$ and $\bar{B}$. 

\paragraph{Discretization} The LRU model is the first of the SSMs that is not seen as a discretization of a continuous-time model. Instead, a discrete parametrization of $\bar A, \ \bar B,\ \bar C, \ \bar D$ is directly used. 

\begin{figure*}[t]
        \centering
        \includegraphics[width=1.\linewidth]{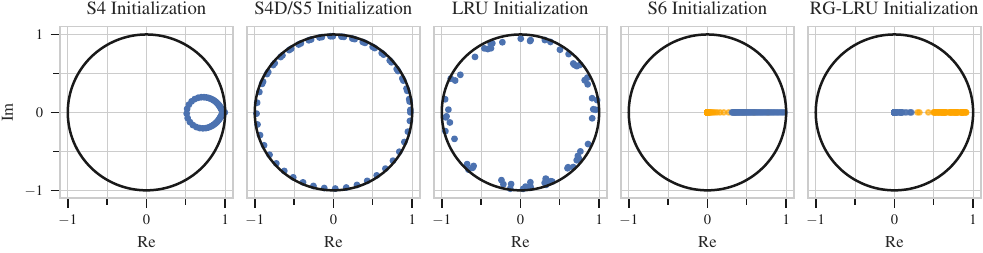}
        \caption{Complex plane representation of the unit disk and the eigenvalues of discrete-time dynamics matrix $\bar A$ \eqref{eqn:dynamics_discrete} resulting from the initialization method in each of the models S4, S4D, S5, LRU, S6, and RG-LRU. Since the initialization of S6 and RG-LRU are input dependent, we plot the initialization for two sample inputs (blue and orange).}
        \label{fig:init}
        \vspace{-12pt}
\end{figure*}

\paragraph{Structure and Initialization} The structure of the model is identical to S5, where a MIMO system --as opposed to $q$ SISO subsystems-- is considered. Given the parametrization~\eqref{eqn: lru_parameterization}, Lemma~\ref{lemm:eigenvalues} is automatically enforced by constraining the eigenvalues of $\bar A$ to lie in the unit-disk. Hence, the initialization is directly performed in polar coordinates by defining a range for $r$ and $\theta$ in which $r$ and $\theta$ are uniformly sampled,  resulting in the eigenvalues shown in Figure~\ref{fig:init}.

\paragraph{Implementation} Similar to LRU, the model is implemented using a parallel scan algorithm~\cite{Blelloch1990} for both training and inference.

\paragraph{Scaffolding} The scaffolding used in LRU is identical to the one used in S5.

\subsection{Scan Selective Structured State Space Sequence Model (S6)}
The S6 parametrization \cite{mamba} introduces for the first time a linear time-varying representation of the dynamics \eqref{eqn:dynamics_continuous}. The time-varying nature of the system stems from the matrices $\bar A_k,\ \bar B_k$ and $\bar C_k$ being functions of the input $u(k)$ at every time-step $k$, which the authors refer to as \emph{selectivity}. Although more expressive, the time-varying representation presents computational challenges. The main contribution of this paper is to address those so the more expressive time-varying nature of the system can be exploited in practice.

\paragraph{Parametrization} Similar to S4D, the S6 parametrization relies on a time-invariant diagonal $A$ matrix \eqref{eqn:s4d_parametrization} as per Assumption \ref{assump:diagonal}. The novelty of the S6 parametrization is that $B$ and $C$ are parameterized to be time-varying given their input-dependent nature: 
\begin{equation}\label{eqn:s6_parametrization}
    B_k = W_B u(k) \qquad C_k = W_C u(k)
\end{equation}
where $W_B$ and $W_C$ are linear projection matrices of appropriate dimensions.

\paragraph{Discretization} Similar to S4D, the S6 model also uses exact discretization to compute the discrete-time dynamics \eqref{eqn:dynamics_discrete}. However, in this case the time-step $\Delta_k$ is itself time-varying since it is a function of the input
\begin{equation}\label{eqn:s6_discretization}
    \Delta_k = \sigma(W_\Delta u(k)), \quad \bar{A}_k = e^{\Delta_k A}, \quad \bar{B}_k = (\Delta_k A)^{-1} (\bar{A}_k - I) \Delta_k B_k,
\end{equation}
$\bar{C}_k = C_k$ and $\bar{D}_k = D_k$, with $W_\Delta \in \mathbb R^{1\times q}$ and $\sigma(\cdot)$ the softplus function.

\paragraph{Structure and Initialization}
Similar to S5, the model is structured in a MIMO manner. In order to initialize the dynamic matrix $A$, its diagonal parametrization is exploited: $\lambda_i = -i\ \forall i = 1, \dots, p$, ensuring that the eigenvalues lie in the negative halfplane. Due to the time-varying nature of the discretization step $\Delta_k$, the eigenvalues of the discrete-time matrices $\bar A_k$ have an initialization that is input-dependent as depicted in Figure~\ref{fig:init}. However, in order to enforce Lemma \ref{lemm:eigenvalues}, the resulting eigenvalues are guaranteed to lie in the unit disk since $\Delta_k$ and $A$ in~\eqref{eqn:s6_discretization} are positive and negative, respectively.

\paragraph{Implementation}
One of the main contributions of the work in \cite{mamba} is to provide an efficient implementation of the time-varying dynamics \eqref{eqn:dynamics_discrete} with matrices \eqref{eqn:s6_parametrization} and \eqref{eqn:s6_discretization} both at inference and training time. In general, the time-varying nature of the S6 model renders the convolutional representation too computationally expensive for practical use. To overcome these limitations, the S6 paper presents a highly customized variation of the parallel scan algorithm~\cite{Blelloch1990} for both training and inference.

\paragraph{Scaffolding}
Another innovation of the work in~\cite{mamba} is the introduction of a new scaffolding: the \emph{Mamba} scaffolding. Here, the pre-processing relies on both linear and nonlinear maps. The map of the upper signal (linear map) is a linear projection followed by a causal convolution, while the map of the lower signal (nonlinear map) is a linear projection followed by a SiLU nonlinearity. The post-processing is once again a gating function similar to previous scaffolding proposals. 

\subsection{Real-Gated Linear Recurrent Unit (RG-LRU)}
The RG-LRU model is a derivative of the well-known long short-term memory (LSTM) model~\cite{Hochreiter1997} and therefore offers a different perspective on SSM models. The RG-LRU model fuses ideas from LSTMs, LRU, and S6.

\paragraph{Parametrization} Following S6, RG-LRU also relies on a time-varying parametrization of the linear dynamics. However, while all previous SSM proposals rely on output feedback dynamics, the RG-LRU model introduces for the first time a state feedback model where $C$ and $D$ are not present. The $A$ and $B$ matrices are then parameterized as
\begin{equation}\label{eqn:rglru_parameterization}
    \bar{A}_k = e^{-c \phi(W_A) \sigma(W_\Delta u(k))}, \qquad \bar{B}_k = \sqrt{1 - A_k^2}\sigma(W_B u(k))
\end{equation}
where $W_\Delta,\ W_A,\ W_B$ are linear projection matrices of appropriate dimensions, $c \in \mathbb{R}$ is a scalar constant,\footnote{The paper empirically found that $c=8$ works best for language modelling.}, $\phi(\cdot)$ is the softplus function, and $\sigma(\cdot)$ is the sigmoid function. The operation $\sqrt{1 - A_k^2}$ is computed element-wise for each entry of $A_k$.

\paragraph{Discretization} Similar to the LRU model, the RG-LRU model does not rely on a continuous-time representation and instead directly parametrizes the discrete matrices $\bar A_k,\ \bar B_k$.

\paragraph{Structure and Initialization} Similar to LRU, the RG-LRU model is structured as a MIMO system. Taking inspiration from LSTMs, this models assumes the state dimension to be equal to the input dimension, i.e., $p=q$. The linear projection matrices $W_\Delta,\ W_A,\ W_B$ are initialized with standard initialization methods, e.g. Glorot~\cite{Glorot2010}, resulting in the eigenvalues shown in Figure~\ref{fig:init}. Given the parameterization of $\bar A_k$ in~\eqref{eqn:rglru_parameterization}, its eigenvalues are restricted to the unit disk by construction.

\paragraph{Implementation} Due to the time-varying nature of the RG-LRU model, it faces the same challenges as the S6 model. Therefore, it also uses a customized variation of the parallel scan algorithm~\cite{Blelloch1990} to compute the outputs at both training and inference time.

\paragraph{Scaffolding}
The RG-LRU model uses the same scaffolding as the S6 model, Mamba. However, this work also introduces two additional task-specific scaffoldings around the basic Mamba scaffolding that are tailored to language modelling: \emph{Hawk} and \emph{Griffin}~\cite[Section~2]{griffin}.

\begin{figure*}
        \centering
        \begin{tabular}{@{}lccccccc@{}}
        \toprule
        \multirow{2}{*}{\textbf{Model}} & \multicolumn{7}{c}{\textbf{LRA Task [\%]}} \\ \cmidrule(lr){2-8}
        & \texttt{ListOps} & \texttt{Text} & \texttt{Retrieval} & \texttt{Image} & \texttt{Pathfinder} & \texttt{Path-X} & avg.\\ \midrule
        Random & 10.00 & 50.00 & 50.00 & 10.00 & 50.00 & 50.00 & 36.67 \\
        Transformer~\cite{Tay2021} (paper results) & 36.37 & 64.27 & 57.46 & 42.44 & 71.40 & FAIL & 53.66\\\midrule
        S4~\cite{Gu2022} (paper results) & 59.60 & 86.82 & 90.90 & 88.65 & 94.20 & 96.35 & 86.09\\
        S4D~\cite{Gu2022s4d} (paper results) & 60.52 & 87.34 & 91.09 & 88.19 & 93.96 & 92.80 & 85.65 \\
        S5~\cite{Smith2023} (paper results) & \textbf{62.15} & 89.31 & \textbf{91.40} & 88.00 & \textbf{95.33} & \textbf{98.58} & \textbf{87.46} \\
        LRU~\cite{Orvieto2023} (paper results) & 60.20 & \textbf{89.40} & 89.90 & \textbf{89.00} & 95.10 & 94.20 & 86.30 \\ \midrule
        S6~\cite{mamba} & 38.02 & 82.98 & 72.14 & 69.82 & 69.26 & 67.32 & 66.59 \\
        RG-LRU~\cite{griffin} & 32.34 & 71.75 & 66.58 & 61.15 & 73.38 & 69.53 & 62.45 \\
        \bottomrule 
        \end{tabular}
        \captionof{table}{Model performance in terms of test accuracy on the LRA benchmark. The first entry (Random) represents the performance of random guessing on the task, i.e., indicating the baseline above which a model is considered to have learned a meaningful representation. Models failing to exceed this baseline on a task are marked as FAIL. The best model on each task is highlighted in \textbf{bold}.}\label{table:lra}
        \vspace{-12pt}
\end{figure*}

\section{Performance in practice} \label{sec:simulations}
In this section, we evaluate the SSM proposals reviewed in Section~\ref{sec:models} on the \emph{long-range arena}~(LRA) benchmark~\cite{Tay2021}. This benchmark evaluates the models in terms of their reasoning ability and their handling of diverse data types, which is one of the strengths of SSMs. We first introduce the benchmark, before presenting the empirical evaluation.

\subsection{Long-Range Arena (LRA) Benchmark}
The goal of the LRA benchmark is to evaluate the reasoning capabilities of sequence models in diverse ways. The benchmark consists of $5$ different tasks, which we summarize in the following. For more details on the benchmark and the individual tasks, we refer to~\cite{Tay2021}.

\paragraph{List Operations (\texttt{ListOps})} This task evaluates a model's ability to capture hierarchical dependencies over long contexts. The goal is to predict the result of a mathematical operation consisting of nested \emph{mean}, \emph{median}, \emph{max}, and \emph{min} operations,\footnote{For instance, 
$\texttt{input: max(4, min(5,6, mean(9, 4, 5)))},$ $\texttt{output:  5}.$}
The task is a ten-way classification task with maximal input lengths of 2k.

\paragraph{Text Classification (\texttt{Text})} This task evaluates a model's ability to capture the tone of long tokenized texts. The dataset consists of IMDb movie reviews, which need to be classified as negative or positive in tone. The task is a binary classification task with maximal input lengths of 4k.

\paragraph{Document Retrieval (\texttt{Retrieval})} This task evaluates a model's ability to compress long sequences into representations that are suitable for similarity matching. The dataset consists of tokenized papers published by the American Academy of Neurology (AAN), which need to be classified in having a citation link or not. The task is a binary classification task with maximal input lengths of 8k.

\paragraph{Image Classification (\texttt{Image})} This task evaluates a model's ability to learn 2D spatial relations from a 1D vector. The dataset consists of vectorized images, which depict one of ten possible classes, e.g. a horse or a car. The task is a ten-way classification task with maximal input lengths of 1k.

\paragraph{Long-Range Spacial Dependency} This task evaluates a model's ability to learn spacial dependencies in a vectorized image. The dataset consists of images, which depict two circles and multiple dashed paths. The goal is to evaluate whether the two circles are connected by any of the present paths or not. The task is therefore a binary classification task and is divided into two subtasks, which only differ in the size of the image. The first subtask has inputs of length 2k and we will refer to it as \texttt{Pathfinder}; the second subtask has a maximal input length of 16k and we will refer to it as \texttt{Path-X}.

\addtolength{\textheight}{-0.5cm}   % This command serves to balance the column lengths

\subsection{Empirical Evaluation of SSM Proposals}
The empirical performance of the reviewed SSM proposals, the Transformer~\cite{Transformer}, and random guessing are reported in Table~\ref{table:lra}. We include the performance of the Transformer as a baseline, since they are the dominant architecture in large language models and sequence modelling. For S4, S4D, S5, and LRU we report the performance of the best variant from the original papers in order to present the most competitive results. Other variants of these models might perform better on tasks not included in the LRA benchmark; for more details on these variants we refer to the original papers. Since performance on the LRA benchmark of S6 and RG-LRU have not been reported in the literature, we provide the results of our own implementation of these architectures, which we make available~\href{https://github.com/jsie7/ssm-benchmark}{here}.\footnote{\url{https://github.com/jsie7/ssm-benchmark}} The hyperparameters of the models and training details of our implementation are stated in the public code repository.

On the LRA benchmark, the LTI-based models S4, S4D, S5, LRU outperform the LTV-based models S6, RG-LRU and the Transformer. From a control theoretic perspective this is surprising, since a general LTV definition encompasses LTI systems as a special case, i.e., a LTV system should perform at least as well as a LTI system. However, this is not the case for the particular time-varying parametrization of S6 or RG-LRU, since e.g.~$\bar A = \bar A_k \ \forall k$ cannot be achieved. We attempted to improve the performance of the LTV-based models by changing the initialization of S6 and RG-LRU and forcing the input-dependent eigenvalues of $\bar A_k \ \forall k$ closer to marginal stability according to Lemma~\ref{lemm:eigenvalues}. However, this resulted in both models to perform considerably worse or fail to learn anything meaningful at all. While marginally stable eigenvalues appear to be important for the LTI-based models, the same is not true for LTV-based models. To date, this behavior is not well understood. Lastly, even though the LTV-based models are closely related to the Transformer~\cite{Ali2024}, they generally perform better on the LRA benchmark.

\section{Conclusion and Future Opportunities} \label{sec:conclusion}
In this paper, we have provided an overview of state-of-the-art state space models (SSM) and explored their features from a control theoretic perspective. In doing this, we highlighted the many connections to standard control theoretic concepts such as the connection between memory and marginal-stability. Additionally, we compared the reviewed SSMs on the long-range arena (LRA) benchmark, finding that the more recent LTV-based SSMs perform worse than their LTI-based counterparts. From a control theoretic perspective, this raises many interesting research questions concerning a LTV parametrization that attains the same performance as the LTI models, as well as a deeper understanding on the role of the eigenvalues in the LTV-based models. 

SSMs, particularly the LTV versions, rely on dynamics where the dynamic matrices depend on the input (excitation) to the system. However, in the SSM literature the theoretical properties arising from these dynamics remains poorly understood. The evident connections between SSMs and linear-system theory give rise to ample opportunities to provide explainability to large foundational models. Moreover, as seen with the LRU model, control theoretic insights have the potential to inform better designs for SSMs.
%Together, the synergy between control-theoretical tools and SSM has the potential to produce the successor of Transformers as the default choice for foundation models.

%%%%%%%%%%%%%%%%%%%%%%%%%%%%%%%%%%%%%%%%%%%%%%%%%%%%%%%%%%%%%%%%%%%%%%%%%%%%%%%%

\bibliography{bibliography.bib}

%%%%%%%%%%%%%%%%%%%%%%%%%%%%%%%%%%%%%%%%%%%%%%%%%%%%%%%%%%%%%%%%%%%%%%%%%%%%%%%%

\end{document}